\documentclass[aip,superscriptaddress]{revtex4-1}

\usepackage{subcaption}
\usepackage{graphicx}
\usepackage{multirow}
\usepackage{amssymb}
\usepackage{amsmath}

\begin{document}


\title{Data-driven Symbolic Closure for Turbulence Modeling in the Lattice Boltzmann Framework} 


\author{Fu Yujie}
\affiliation{School of Mathematics and Computer Sciences, Nanchang University, Nanchang, Jiang Xi, 330031, China}
\affiliation{Institute of Mathematics and interdisciplinary Sciences, Nanchang University, Nanchang, Jiang Xi, 330031, China}

\author{Zhang Yihan}
\affiliation{School of Mathematics and Computer Sciences, Nanchang University, Nanchang, Jiang Xi, 330031, China}
\affiliation{Institute of Mathematics and interdisciplinary Sciences, Nanchang University, Nanchang, Jiang Xi, 330031, China}

\author{Deng Wanru}
\affiliation{School of Civil Engineering, Tianjin University, Tianjin, 300350, China}
\affiliation{National Science Center for Earthquake Engineering, Tianjin University, Tianjin, 300350, China}

\author{Dai Yuanjun}
\email[Corresponding Author: ]
{iankin.dai@ncu.edu.cn}
\affiliation{School of Mathematics and Computer Sciences, Nanchang University, Nanchang, Jiang Xi, 330031, China}
\affiliation{Institute of Mathematics and interdisciplinary Sciences, Nanchang University, Nanchang, Jiang Xi, 330031, China}


\date{\today}

\begin{abstract}
Turbulence modeling within the Lattice Boltzmann Method (LBM) framework has long relied on traditional algebraic sub-grid scale (SGS) models, which often suffer from over-dissipation and lack of spatial selectivity near solid boundaries. In this work, we utilize Physical Symbolic Optimization ($\Phi$-SO) to discover explicit analytical closures from high-fidelity DNS datasets of Taylor-Green Vortex (TGV) and Lid-Driven Cavity (LDC) flows. Central to our methodology is the integration of virtual dimensional analysis and non-linear tensor invariants—a strategy that enforces physical scaling laws directly within the symbolic search process. The resulting model exhibits a highly non-linear dependency on both strain-rate and rotation-rate invariants. Numerical validations confirm that this symbolic closure outperforms the standard Smagorinsky approach in capturing kinetic energy dissipation rate peaks and resolving delicate secondary corner vortices. Furthermore, the model exhibits robust zero-shot generalization to wall-bounded turbulent channel flow ($Re_{\tau} = 180$) without the aid of any supplemental wall-damping corrections. This work highlights the potential of symbolic regression to uncover robust, interpretable physical laws for the next generation of intelligent computational fluid dynamics solvers.
\end{abstract}

\pacs{47.27.ep, 47.27.Gh, 47.11.-j}

\keywords{symbolic regression, lattice Boltzmann method, large eddy simulation, turbulence modeling, subgrid-scale model}

\maketitle 

\section{Introduction}
Turbulence remains one of the most challenging and open problems in classical physics and engineering \cite{Pope2000}. The chaotic, multi-scale nature of turbulent flows demands numerical methods capable of capturing intricate flow structures across a wide range of spatial and temporal scales. In Computational Fluid Dynamics (CFD), three primary methodologies have been established: Direct Numerical Simulation (DNS), Reynolds-Averaged Navier-Stokes (RANS), and Large Eddy Simulation (LES) \cite{Ferziger_2002}. DNS offers the highest fidelity by resolving all turbulent scales, but its computational cost renders it prohibitive for high-Reynolds-number engineering applications \cite{Moin_1998}. Conversely, RANS, though computationally efficient, averages out transient fluctuations, limiting its accuracy in unsteady, detached flows \cite{Spalart_2000}. LES strikes a balance by resolving large-scale energy-containing eddies while modeling the isotropic, universal small-scale motions through Sub-Grid Scale (SGS) models \cite{smagorinsky1963,Sagault}. Consequently, LES has become the preferred approach for balancing fidelity and efficiency in turbulence research \cite{Yang2015}.

Parallel to the advancement of LES, the Lattice Boltzmann Method (LBM) has emerged as a robust alternative to traditional Navier-Stokes solvers based on Finite Volume (FVM) or Finite Difference Methods (FDM). Rooted in mesoscopic kinetic theory, LBM models fluid dynamics through the streaming and collision of particle distribution functions. The inherent locality of data operations makes it exceptionally amenable to massive parallelization, a critical advantage in the era of GPU-accelerated high-performance computing. Furthermore, LBM's straightforward boundary handling and algorithmic simplicity make it particularly promising for simulating complex geometries and transient flows. The integration of LES within the LBM framework has been extensively studied, leading to the development of various turbulence closure models \cite{YU2006957,Xia02012015}. The core challenge in LBM-LES lies in determining the effective relaxation time, which corresponds to the eddy viscosity in the macroscopic limit. The earliest and most widely used approach is the Smagorinsky model \cite{smagorinsky1963}, proposed in 1963. In this model, the eddy viscosity is assumed to be linearly related to the magnitude of the local strain-rate tensor: $\nu_t = (C_s \Delta)^2 |S|$, where $C_s$ is the Smagorinsky constant (typically $C_s = 0.1$--$0.2$). In this study, we use $C_s = 0.1$ for all Smagorinsky simulations. While the Smagorinsky model yields satisfactory results in homogeneous isotropic turbulence, it suffers from critical limitations in wall-bounded flows. Specifically, it is known to be over-dissipative near solid walls \cite{FREIRE2021105135} and introduces non-physical eddy viscosity in laminar shear regions \cite{10.1063/1.4882880}, inaccuracies that are evident in velocity profiles and energy spectra \cite{10.1063/1.1839152}.To mitigate these deficiencies, researchers have proposed several advanced algebraic closures. Nicoud and Ducros \cite{Nicoud1999} introduced the Wall-Adapting Local Eddy-viscosity (WALE) model, which scales the eddy viscosity using the square of the velocity gradient tensor. This modification allows the model to correctly capture the asymptotic near-wall behavior of turbulent dissipation. Similarly, Vreman et al. \cite{Vreman2004} developed an eddy viscosity model based on the algebraic relationship between velocity gradient tensor components, effectively eliminating non-physical viscosity in laminar flows. Other notable contributions include the $\sigma$-model \cite{Nicoud2011}, based on the singular values of the velocity gradient tensor, and the Volumetric Strain-Stretching (VSS) model \cite{10.1063/1.4882880}.Despite these improvements, existing algebraic models still face inherent limitations. For instance, theoretical analyses by Nicoud et al. \cite{Nicoud1999,Nicoud2011} demonstrated that the eddy viscosity in the Vreman model decays linearly with respect to wall distance, which deviates from the physically correct cubic behavior required in the near-wall region. Moreover, most of these models rely on fixed theoretical assumptions or empirical constants that may not be universally applicable across diverse flow regimes. In recent years, the development of new purely algebraic SGS models appears to have reached a bottleneck, with few breakthrough theories emerging. This stagnation suggests that traditional derivation methods may be approaching their limits, necessitating a paradigm shift towards new discovery methodologies.

In recent years, data-driven turbulence modeling has gained significant traction \cite{Duraisamy2019}. Machine learning (ML) techniques have demonstrated remarkable success in mapping flow features to SGS stresses or turbulent viscosities, often outperforming traditional algebraic models \cite{Maulik_San_Rasheed_Vedula_2019,Taghizadeh2021}. For example, recent works have utilized Convolutional Neural Networks (CNNs) and Generative Adversarial Networks (GANs) to reconstruct super-resolution flow fields or predict Reynolds stresses \cite{Fukami_Fukagata_Taira_2019,PhysRevFluids.6.124602}. Despite their high accuracy, neural network-based models suffer from a "black-box" nature. They lack physical interpretability and are difficult to implement in existing CFD solvers without coupling to heavy ML inference libraries. This opacity hinders the extraction of physical insights and limits the trust in engineering applications. 

To bridge the gap between data accuracy and interpretability, Symbolic Regression (SR) has emerged as a promising alternative. Unlike fixed-structure regression or opaque neural networks, SR searches the space of mathematical expressions to discover analytical equations that best describe the data. Pioneering works by Brunton et al. \cite{Steven2016} and Schmelzer et al. \cite{Schmelzer2019} have successfully applied sparse regression to discover algebraic Reynolds Stress models for RANS. More recently, highly advanced evolutionary algorithms and Deep Symbolic Optimization methods have been applied to turbulence modeling. For instance, Liao et al. (2025) proposed a mutually coupled framework of SR and data assimilation for airfoil flows \cite{Hui2025}, and Reissmann et al. (2025) utilized language model-based transfer learning to accelerate symbolic model discovery for complex geometries \cite{Reissmann2025}. These studies, primarily in the Navier-Stokes framework, demonstrate that SR can yield parsimonious, physically interpretable models that generalize well.

However, the application of symbolic regression specifically for LBM-based LES closures remains relatively unexplored. The unique mesoscopic formulation of LBM (distribution functions, relaxation time) requires tailored closure strategies different from macroscopic RANS/LES stress tensors. In this study, we propose a data-driven framework to discover explicit algebraic closures for $\tau_{\mathrm{eff}}$ within the LBM framework. We utilize the Physical Symbolic Optimization ($\Phi$-SO) algorithm \cite{Tenachi_2023}, integrating high-fidelity DNS data from Taylor-Green Vortex (TGV) and Lid-Driven Cavity (LDC) flows. The distinct contributions of this work are:
\begin{itemize}
    \item Physics-Informed Constraints: We introduce virtual dimensional analysis and non-linear tensor invariants (e.g., coupling rotation and strain) into the symbolic search, ensuring the discovered model respects physical scaling laws.
    \item Interpretable \& Explicit: The resulting model is a compact algebraic equation that can be directly implemented in open-source solvers like \textbf{OpenLB} \cite{olbRelease17} without external ML dependencies.
    \item Zero-Shot Generalization: We demonstrate that the model, trained only on simple canonical flows, successfully generalizes to wall-bounded Turbulent Channel Flow ($Re_{\tau}=180$) without needing ad-hoc wall damping functions, outperforming the classic Smagorinsky model.
\end{itemize}

The remainder of this paper is organized as follows: Sec. \ref{sec:methodology} presents the computational framework, covering the generation of high-fidelity DNS datasets (TGV and LDC), the extraction of the effective relaxation time via spatial filtering, and the configuration of the $\Phi$-SO with virtual dimensional constraints. Subsequently, the discovered algebraic model is explicitly formulated and rigorously validated in Sec. \ref{sec:results}. This validation includes a comprehensive comparative analysis against classical LES models across three distinct flow regimes: decaying homogeneous isotropic turbulence, inhomogeneous shear flow with recirculation, and a generalization test on wall-bounded turbulent channel flow. Finally, the conclusions and potential avenues for future research are summarized in Sec. \ref{sec:conclusion}.

\section{Methodology}
\label{sec:methodology}
\begin{figure}[ht]
    \centering
    \begin{subfigure}{\linewidth}
        \includegraphics[width=\linewidth]{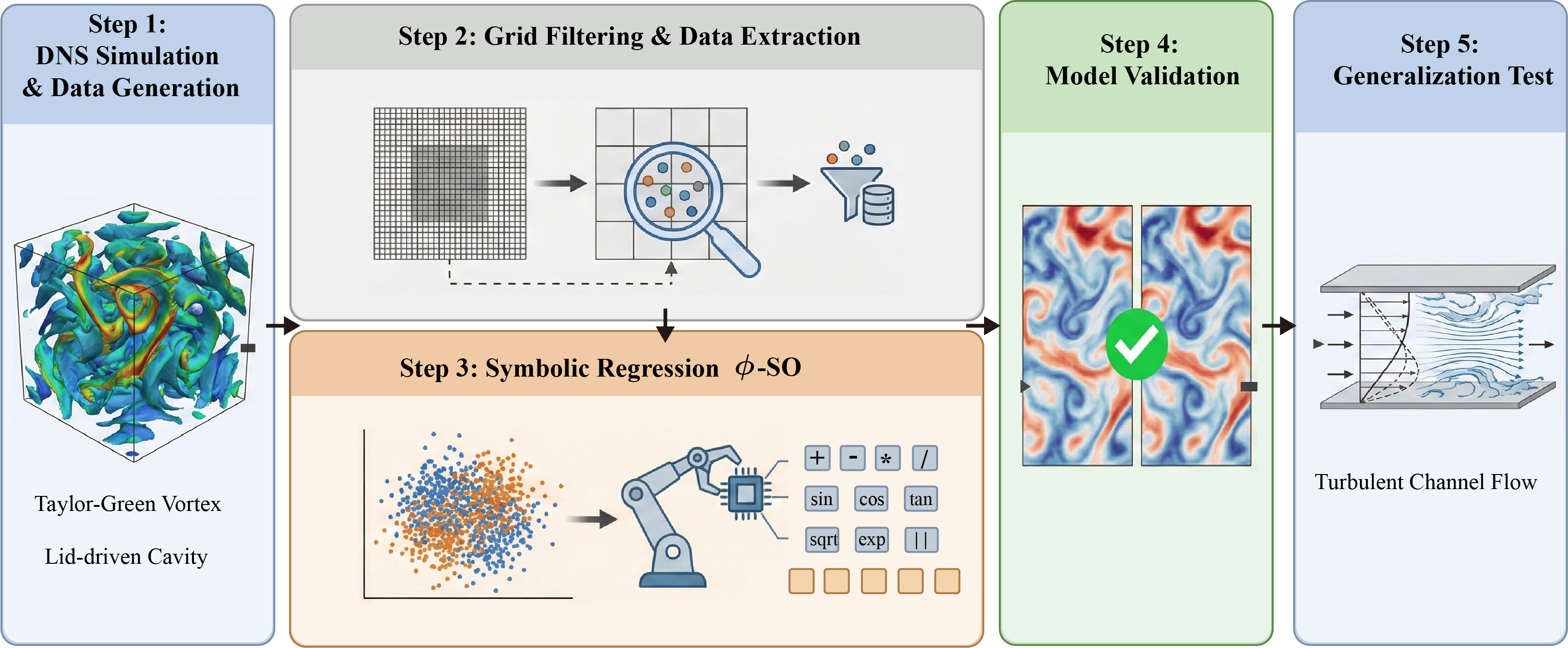}
        \caption{Workflow}
        \label{fig:1-a}
    \end{subfigure}
    \quad
    \begin{subfigure}{\linewidth}
        \includegraphics[width=\linewidth]{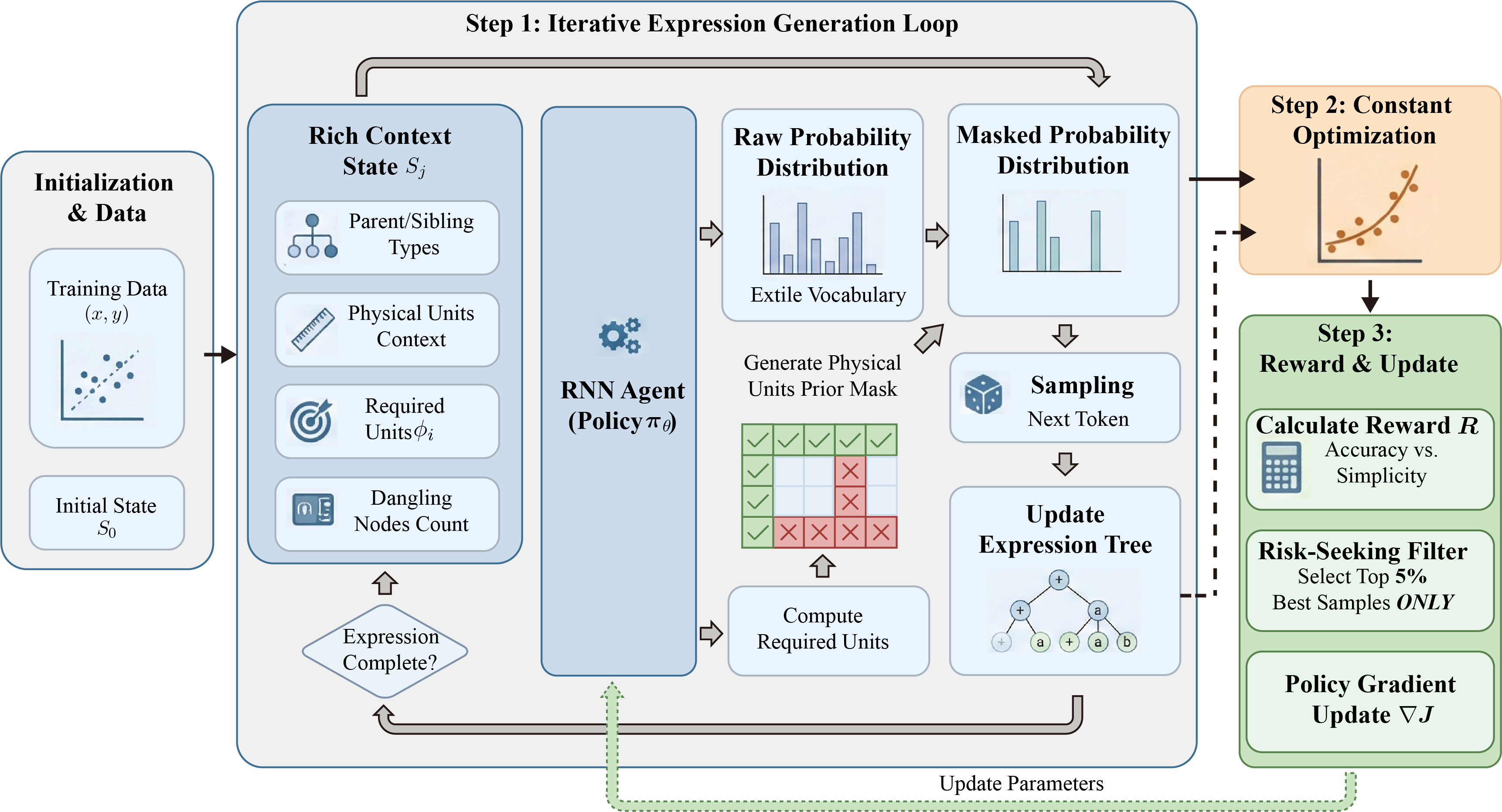}
        \caption{$\Phi$-SO Architecture}
        \label{fig:1-b}
    \end{subfigure}
    \caption{Schematic illustration of data-driven symbolic closure for turbulence modeling}
    \label{fig:placeholder}
\end{figure}

As outlined in the introduction, this study aims to discover an interpretable, data-driven turbulence model within the LBM framework. The proposed workflow integrates high-fidelity DNS data with a physical symbolic optimization algorithm to derive an explicit algebraic formulation for the effective relaxation time, $\tau_{\mathrm{eff}}$. The overall workflow, from data generation to model validation, is illustrated in Fig.~\ref{fig:1-a}.

\subsection{High-Fidelity Data Generation}
The cornerstone of this data-driven discovery framework is a high-fidelity dataset generated by DNS, which resolves all relevant scales of turbulent motion. The DNS data serves a dual purpose: (\romannumeral1) to provide the ground truth for training the symbolic regression model, and (\romannumeral2) to establish a benchmark for the final a posteriori validation of the discovered closure model. Two canonical flow configurations, TGV and LDC flow, are simulated to capture different aspects of turbulence physics. All simulations are performed using the well-validated, open-source Lattice Boltzmann code \textbf{OpenLB} \cite{olbRelease17}.

\textbf{1. TGV: Homogeneous Isotropic Turbulence}

The TGV is a fundamental benchmark for studying the transition to turbulence and subsequent decay of homogeneous isotropic turbulence. Our DNS configuration follows the classic setup of Brachet et al. \cite{Brachet_Meiron_Orszag_Nickel_Morf_Frisch_1983}. A three-dimensional periodic domain of size $\left[0,2\pi\right]^3$ is discretized using a uniform lattice with $N^3=256^3$ cells. The simulation employs the D3Q19 lattice and the single-relaxation-time (BGK) collision model \cite{PhysRevA.45.R5339,Y.H.Qian_1992}. The key control parameter, the Reynolds number ($Re=V_0L/\nu$), is varied by adjusting the kinematic viscosity $\nu$ to target $Re=800,1600$, and $3000$. The flow evolution is simulated until the final dimensionless time $t=10(L/V_0)$, capturing the initial vortex stretching, breakdown, and turbulence decay phases. The temporal evolution of the volume-averaged kinetic energy for all three cases shows excellent agreement with the spectral method results of Brachet et al. \cite{Brachet_Meiron_Orszag_Nickel_Morf_Frisch_1983}, which validates the accuracy of our DNS dataset for homogeneous, decaying turbulence.

\textbf{2. LDC Flow: Inhomogeneous Wall-Bounded Turbulence}

To incorporate shear and wall effects absent in TGV, DNS of a three-dimensional LDC is conducted. The flow in a cubic cavity of side length $L=1$ is driven by the steady motion of the top lid with a constant velocity $U_{\text{lid}}=0.1$ in the x-direction. No-slip conditions are applied on all walls. A uniform grid of $256^3$ cells is used, and the Reynolds number is defined as $Re=U_{\text{lid}}L/\nu$.

The accuracy of our LDC DNS data is validated against the seminal experimental work of Prasad and Koseff \cite{PrasadAndKoseff}, which provides comprehensive measurements for 3D cavity flows. We simulated two turbulent regimes at $Re=3200$ and $Re=10,000$, matching the experimental conditions. The simulations were advanced until a statistically steady state was achieved, followed by a prolonged period of time-averaging to acquire mean flow statistics. A comparison confirms that our DNS results are in excellent agreement with the experimental data across key turbulent statistics, including mean velocity profiles and Reynolds stress components. 
\subsection{LBM-LES Filtering and Data Extraction}
\label{sec:3-2}
The objective of this step is to create datasets for symbolic regression. The workflow begins with the high-resolution DNS results obtained from \textbf{OpenLB}, which provide the complete distribution functions \( f_i(\mathbf{x}, t) \) and their equilibrium counterparts \( f_i^{\mathrm{eq}}(\mathbf{x}, t) \) at each lattice node and time step. To emulate an LES, a spatial filter is applied to these distribution functions. The filter width \( \Delta \) is chosen to correspond to the intended coarse grid resolution of the LES. We employ a Gaussian filter kernel \( G \) for this purpose. The filtered distribution functions are computed as follows:
 \begin{eqnarray}
     \bar{f}_i(\mathbf{x}_{\mathrm{LES}}, t) &= G * f_i(\mathbf{x}_{\mathrm{DNS}}, t)\\
     \bar{f}_i^{\mathrm{eq}}(\mathbf{x}_{\mathrm{LES}}, t) &= G * f_i^{\mathrm{eq}}(\mathbf{x}_{\mathrm{DNS}}, t)
 \end{eqnarray}
 where \( * \) denotes the convolution operation, and \( \mathbf{x}_{\mathrm{LES}} \) represents the coordinates on the coarser LES grid.
For the BGK collision model, the filtered LBM equation takes the form:

\begin{equation}
\bar{f}_i(\mathbf{x} + \mathbf{c}_i \Delta t, t + \Delta t) - \bar{f}_i(\mathbf{x}, t) = -\frac{1}{\tau_{\mathrm{eff}}} \left[ \bar{f}_i(\mathbf{x}, t) - \bar{f}_i^{\mathrm{eq}}(\mathbf{x}, t) \right]
\end{equation}
From this equation, the effective relaxation time \( \tau_{\mathrm{eff}} \), which encapsulates the combined effects of molecular viscosity and unresolved turbulent scales, can be determined. Due to the enormous number of data points (considering all spatial nodes and time steps), a random sampling strategy is adopted to ensure computational tractability. To mitigate the influence of numerical noise, a least-squares approach is used to solve for \( \tau_{\mathrm{eff}} \) at each sampled point:

\begin{equation}
\tau_{\mathrm{eff}} = -\frac{ \sum_i \left[ \bar{f}_i(\mathbf{x}, t) - \bar{f}_i^{\mathrm{eq}}(\mathbf{x}, t) \right]^2 }{ \sum_i \left[ \bar{f}_i(\mathbf{x}, t) - \bar{f}_i^{\mathrm{eq}}(\mathbf{x}, t) \right] \left[ \bar{f}_i(\mathbf{x} + \mathbf{c}_i \Delta t, t + \Delta t) - \bar{f}_i(\mathbf{x}, t) \right] }
\end{equation}

Points where the denominator approaches zero (indicating numerical instability or negligible collision effect) are excluded from the dataset. The turbulent contribution to the relaxation time, denoted \( \tau_{\mathrm{turb}} \), is then obtained by subtracting the known molecular relaxation time \( \tau_0 \) (associated with the laminar viscosity):

\begin{equation}
\tau_{\mathrm{turb}} = \tau_{\mathrm{eff}} - \tau_0
\end{equation}

This quantity \( \tau_{\mathrm{turb}} \) serves as the target variable that the symbolic regression model will learn to predict based on local flow features.

The input features for the model are derived from the filtered velocity gradient tensor \( \mathbf{A}_{ij}=\partial_j\mathbf{\bar{u}}_i \). A set of frame-indifferent invariants, commonly used in turbulence modeling for their physical significance, are computed. These include the strain-rate tensor  \( \mathbf{S}_{ij} = \frac{1}{2}(\mathbf{A}_{ij} + \mathbf{A}_{ji}) \), the rotation-rate tensor \( \Omega_{ij} = \frac{1}{2}(\mathbf{A}_{ij} - \mathbf{A}_{ji}) \), along with their basic and higher-order invariants ($I$,$II$,$III$,$IV$).

For each sampled data point, a feature vector \( \mathbf{X} \) is constructed from these quantities, and the target variable \( y \) is defined as the turbulent relaxation time increment. The specific composition is:

\begin{itemize}
    \item \textbf{Feature vector}: \( \mathbf{X} = [|S|, |\Omega|, I_S, I_{\Omega}, II_S, II_{\Omega}, III_S, IV_{S\Omega}] \)
    \item \textbf{Target variable}: \( y = \tau_{\mathrm{turb}} \)
\end{itemize}

This comprehensive set of features is designed to provide the symbolic regression algorithm with a rich description of the local flow state, encompassing strain, rotation, dissipation, and near-wall effects, thereby enabling the discovery of a physically meaningful closure model. The resulting dataset, comprising pairs of \( (\mathbf{X}, y) \), forms the direct input to the symbolic optimization process described in the next section.
\subsection{Symbolic Regression via $\Phi$-SO}
\label{sec:physo}
The core of the data-driven closure model discovery is performed using the $\Phi$-SO framework, a state-of-the-art symbolic regression algorithm \cite{Tenachi_2023}. The algorithm operates in an iterative, closed-loop manner, as illustrated in Fig.~\ref{fig:1-b}. Its objective is to identify the optimal symbolic expression $\mathcal{N}( \mathbf{X} ) $ that maps the input feature vector \( \mathbf{X} \) to the target variable \(\tau_{\mathrm{turb}} \) with high fidelity while maintaining a simple form.

\begin{itemize}
    \item \textbf{Step 1: Iterative Expression Generation Loop.} The process begins with the initialization of a recurrent neural network (RNN) agent, which serves as a policy $\pi_{\theta}$ for generating mathematical expressions. Starting from an initial state $S_0$, the agent constructs an expression tree token by token. At each step $j$, the RNN considers a rich context state $S_j$, which includes information about the partially built expression. Based on this context, the agent outputs a probability distribution over possible next tokens. A next token is sampled from this distribution, and the expression tree is updated. This loop continues until a complete, syntactically valid expression is generated. The use of an RNN allows the algorithm to efficiently explore the vast combinatorial space of possible expressions.
    \item \textbf{Step 2: Constant Optimization.} Once a candidate expression structure is generated, its continuous parameters (e.g., multiplicative coefficients, additive constants) are optimized. This is achieved by treating the symbolic expression as a function with tunable constants and minimizing the mean squared error (MSE) between its predictions and the target data \(\tau_{\mathrm{turb}} \) using a gradient-based optimizer. This step ensures that for a given expression form, the constants are set to their optimal values, providing a fair evaluation of the expression's inherent capacity to fit the data.
    \item \textbf{Step 3: Reward Calculation and Policy Update.} Each candidate expression, with its constants optimized, is then evaluated using a reward function $R$. The reward balances two critical aspects: accuracy and simplicity. Accuracy is measured by the goodness-of-fit to the data, while simplicity is measured by the expression's complexity. The specific reward function used is:
    \begin{equation}
        R=-\mathrm{MSE}-\lambda\cdot\mathrm{Complexity},
    \end{equation}
    where $\lambda$ is a regularization parameter that controls the trade-off. Following a risk-seeking policy gradient strategy, only the top-performing candidates are used to compute the gradient $\nabla J$ for updating the RNN policy parameters $\theta$. This reinforcement learning approach progressively biases the expression generator towards discovering higher-reward, and thus more accurate and simpler, symbolic expressions over many iterations.
\end{itemize}

In this work, the $\Phi$-SO algorithm is applied to the dataset prepared in Section \ref{sec:3-2}. The search space of mathematical operations includes basic arithmetic, exponents, and elementary functions. The algorithm performs multiple independent runs to explore the solution space thoroughly. From the Pareto front of discovered expressions—which represents the trade-off between accuracy and complexity—we select the most promising candidate for further analysis and integration into the LBM solver. This selected model, denoted as \(\tau_{\mathrm{turb}} =\mathcal{N}(\mathbf{X})\), constitutes the novel, data-driven turbulence closure.
\subsection{Implementation and Validation}
The optimal expression identified by $\Phi$-SO was implemented directly into the \textbf{OpenLB} open-source software package:
\begin{itemize}
    \item \textbf{Validation:} The model was first tested on the training geometries (TGV and LDC) to verify that it could reproduce the statistical features of the DNS data and outperform classic SGS models like Smagorinsky.
    \item \textbf{Generalization:} To assess the model's universality, we simulated a Turbulent Channel Flow---a scenario absent from the training phase. The results (velocity profiles and Reynolds stresses) were compared against standard benchmark data to demonstrate the model's ability to generalize to new wall-bounded flow regimes.
\end{itemize}
\section{Results}
\label{sec:results}
\subsection{Discovery of the Symbolic Closure Model}
The discovery of the turbulence closure model was conducted using the $\Phi$-SO framework with a configuration specifically tailored for Lattice Boltzmann variables. While the core algorithm follows the standard setup described in Section \ref{sec:physo}, we introduced specific physics-informed constraints to guide the search process towards physically meaningful solutions. 
Initially, the symbolic regression was performed using only the raw local strain rate ($S_{ij}$) and rotation rate ($\Omega_{ij}$) tensors along with their basic invariants. However, preliminary tests indicated that these basic features were insufficient to capture the complex anisotropic behavior of the turbulent viscosity, yielding suboptimal correlation with the DNS data. To address this, we enhanced the feature space by explicitly constructing higher-order non-linear tensor invariants, such as terms derived from $\Omega^2$, $\Omega S$, and their coupled products. Furthermore, although the LBM simulation is conducted in lattice units (dimensionless), we introduced a virtual dimensional analysis strategy to embed prior physical knowledge. Based on the analysis of the eddy viscosity scaling, we imposed a constraint that the effective relaxation time $\tau_{\mathrm{eff}}$ should exhibit a scaling behavior proportional to the strain magnitude ($|S|$). To enforce this, we assigned virtual physical dimensions to the dimensionless LBM variables in the $\Phi$-SO library. This allowed the algorithm's dimensional verification module to prune the search space of physically inconsistent expressions. Crucially, our dimensional analysis revealed that the constructed non-linear features could reach virtual dimensions as high as order $S^4$ (e.g., in terms like $II_{S}^2$). To allow the regression model to map these high-order terms back to the required target dimensionality (order $S^1$), it was essential to introduce fractional power operators. Therefore, we explicitly added the pow(1/4) operator to the function library. This addition proved to be a critical factor in discovering the final converged model.

The $\Phi$-SO algorithm generated a Pareto front of candidate solutions, representing the optimal trade-off between expression complexity and prediction accuracy. We selected the model located at the "knee point" of the Pareto front, which offers the best balance—minimizing error while avoiding the excessive complexity that leads to overfitting. The discovered symbolic closure model for the effective relaxation time is given by:

\begin{equation}
\label{eq:model_tau}
\tau_{\mathrm{turb}} = 0.26 \Delta^2 \frac{A \cdot B^{0.5}}{D}
\end{equation}
where the intermediate terms $A$, $B$, and the denominator $D$ are formulated as:
\begin{align}
A &= 1.03 II_{\Omega}^2 + 7.95 II_{\Omega} II_{S} + 1.05 II_{S}^2 + 2.87 IV_{S\Omega} \label{eq:A}\\
B &= A \left( 1 + 0.04 \frac{II_{\Omega}}{II_{S}} \right) \label{eq:B}\\
D &= 2.71 II_{\Omega}^2 A^{0.25} + 21.55 II_{\Omega} II_{S} A^{0.25} + 2.69 II_{S}^2 A^{0.25} + 25.46 C \label{eq:D}
\end{align}
with $C$ representing a high-order correction term:
\begin{equation}
C = (-II_{S})^2 (-II_{S} + 0.27 I_{S})^{0.5} \label{eq:C}
\end{equation}

The discovered formula reveals a highly non-linear dependency on both the strain rate and rotation rate invariants ($II_S$, $II_{\Omega}$). Unlike the standard Smagorinsky model, which relies solely on the strain rate magnitude ($|S|$), Eq. (\ref{eq:model_tau}) incorporates the interaction between rotation and strain (through terms like $II_{\Omega} II_{S}$ and $IV_{S\Omega}$), suggesting that the model can inherently distinguish between laminar shear and turbulent rotation. \emph{It is worth noting that while the model demonstrates excellent empirical performance (as detailed in the subsequent sections), the precise theoretical derivation of the specific fractional exponents and coefficients remains complex.} The current formulation is a result of data-driven discovery rather than first-principles derivation. A rigorous theoretical analysis to explicitly map these data-learned terms to classical turbulence theory is limited by our current theoretical framework and remains an open topic for future research. Nevertheless, the results indicate that this algebraic structure effectively captures the sub-grid scale physics required for LBM-LES.

\subsection{A posteriori Validation: Homogeneous Isotropic Turbulence}
To assess the physical fidelity of the discovered symbolic closure, we first consider the TGV flow. This case serves as a rigorous benchmark for verifying the model's ability to capture the transition from large-scale ordered laminar flow to small-scale disordered turbulence.

\begin{figure}[ht]
    \centering
    \begin{subfigure}[b]{\linewidth}
        \centering
        \includegraphics[width=\linewidth]{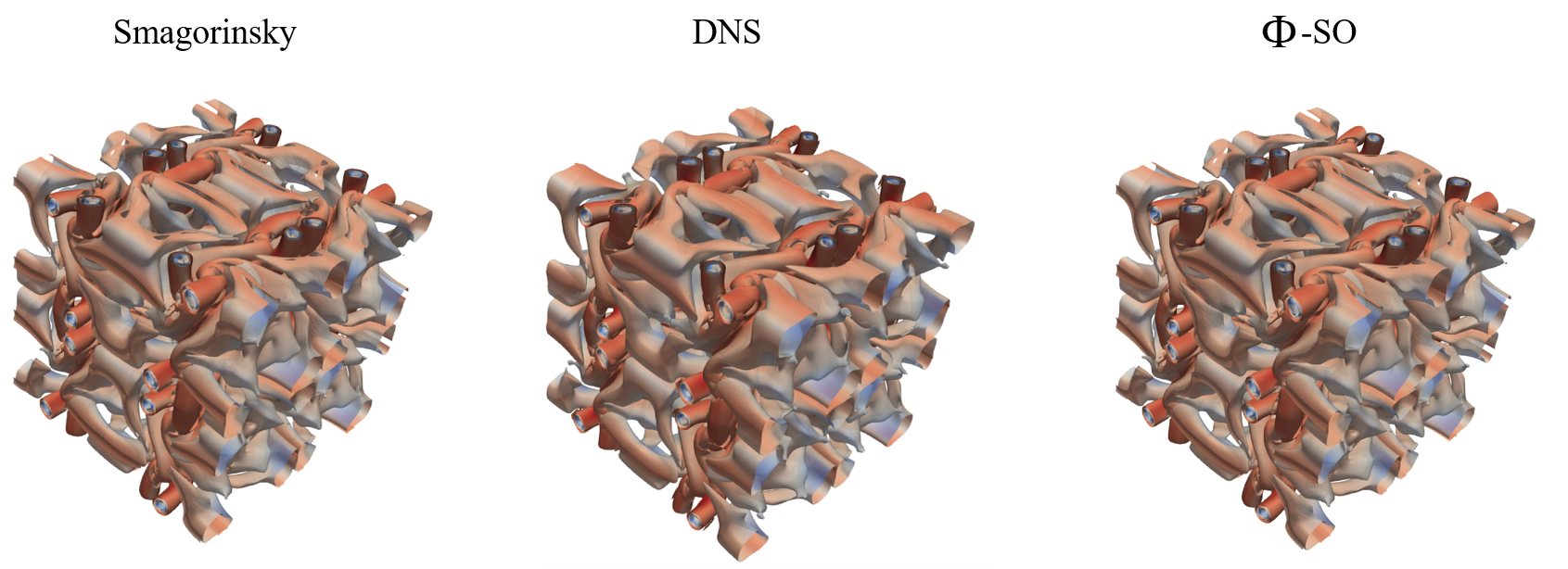}
        \caption{$t=6$}
        \label{fig:tgv_flow_strcuture_t6}
    \end{subfigure}
    \begin{subfigure}[b]{\linewidth}
        \centering
        \includegraphics[width=\linewidth]{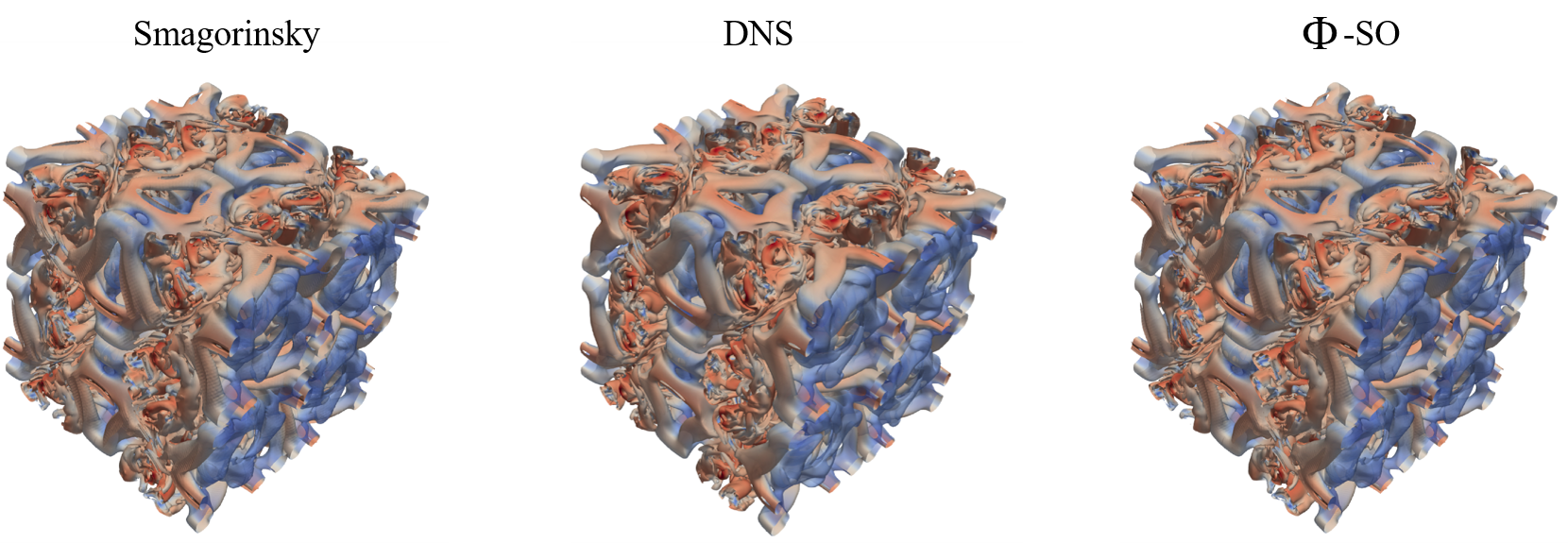}
        \caption{$t=9$}
        \label{fig:tgv_flow_strcuture_t9}
    \end{subfigure}
    \caption{Comparison of flow structures between LBM-DNS, LBM-Smagorinsky, and LBM-$\Phi$-SO at  $Re=1600$. All models captured the characteristics of TGV, that is the large vortices breakdown into smaller vortices then cascaded into even smaller ones.}
    \label{fig:tgv_flow_strcuture}
\end{figure}

Figure \ref{fig:tgv_flow_strcuture} presents the qualitative comparison of instantaneous flow structures (visualized via Q-criterion iso-surfaces) at $Re=1600$. The snapshots correspond to the vortex breakdown phase ($t=6$) and the developed turbulent phase ($t=9$). From a visual perspective, all three simulations—Filtered DNS, standard Smagorinsky, and the $\Phi$-SO data-driven model—successfully reproduce the characteristic phenomenology of the TGV cascade. As shown in the figure, the initial large-scale ordered vortices break down into smaller coherent structures, eventually forming a cloud of isotropic small-scale eddies. This qualitative agreement demonstrates that the discovered symbolic model is capable of sustaining the fundamental vortex dynamics expected in turbulent flows, performing at least on par with established LES sub-grid models.

\begin{figure}[t]
    \centering
    \begin{subfigure}[b]{0.32\linewidth}
        \centering
        \includegraphics[width=1.15\linewidth]{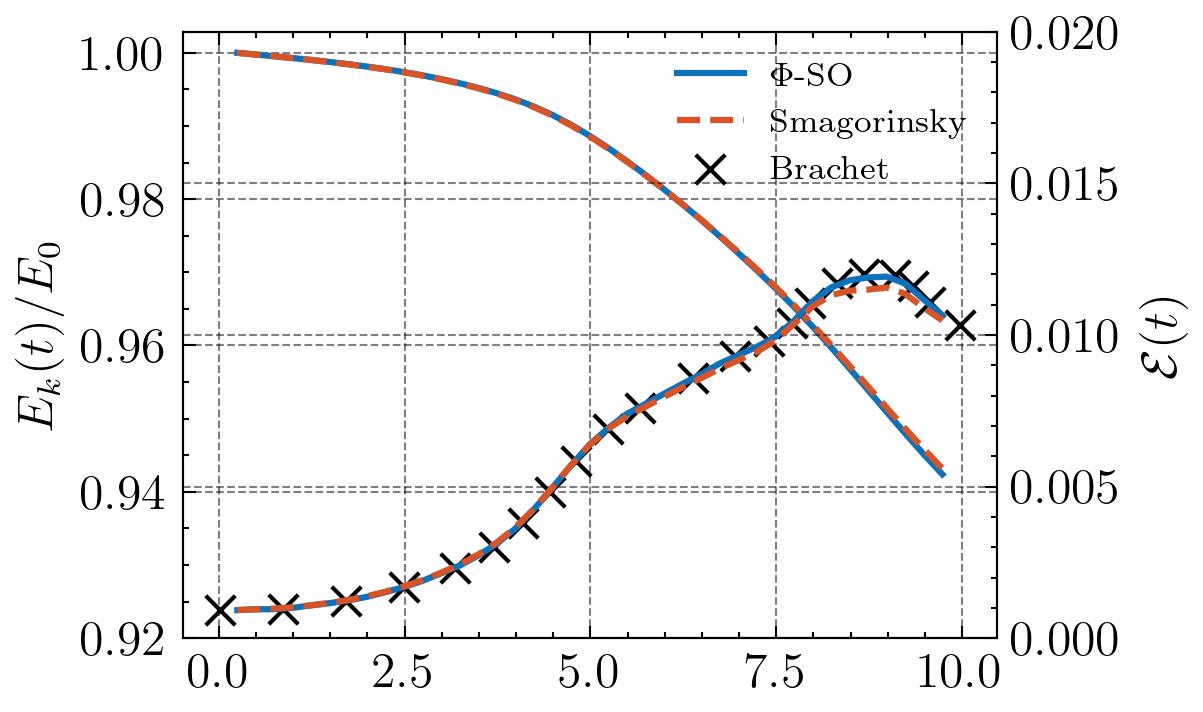}
        \caption{$Re=800$}
        \label{fig:tgv_re800}
    \end{subfigure}
    \hfill
    \begin{subfigure}[b]{0.32\linewidth}
        \centering
        \includegraphics[width=1.15\linewidth]{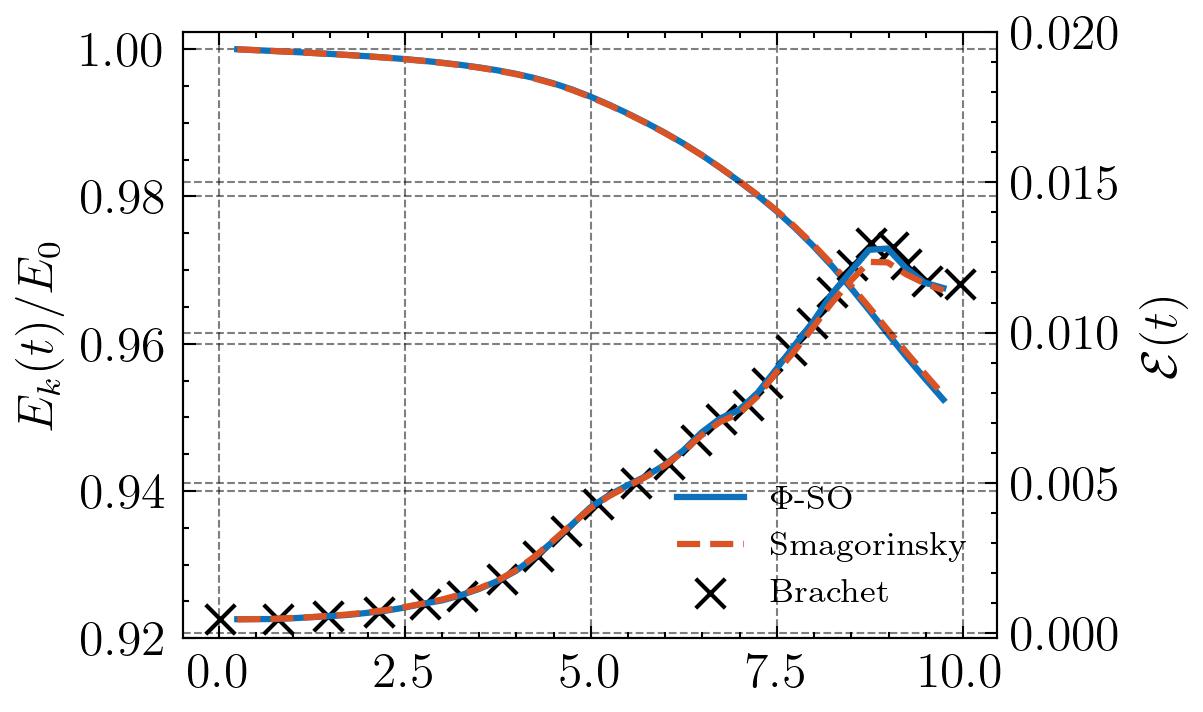}
        \caption{$Re=1600$}
        \label{fig:tgv_re1600}
    \end{subfigure}
    \hfill
    \begin{subfigure}[b]{0.32\linewidth}
        \centering
        \includegraphics[width=1.15\linewidth]{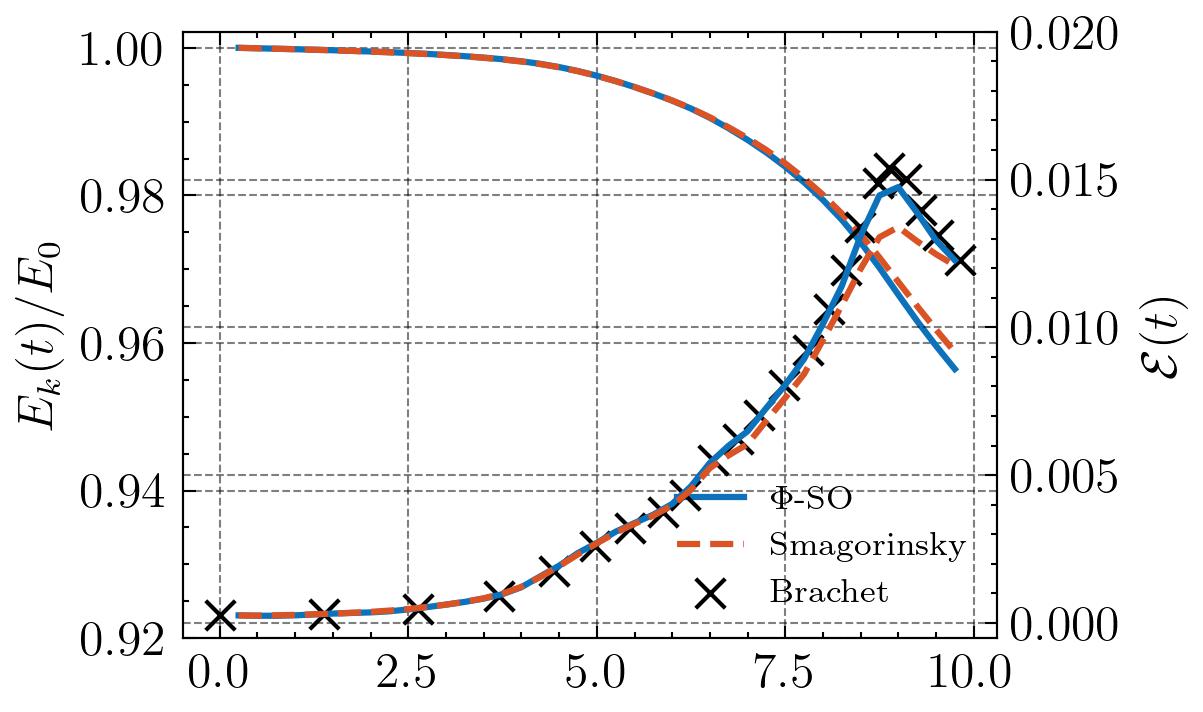}
        \caption{$Re=3000$}
        \label{fig:tgv_re3000}
    \end{subfigure}
    
    \caption{Temporal evolution of normalized total kinetic energy $E_k(t)/E_0$ and dissipation $\mathcal{E}(t)$ for the Taylor-Green Vortex at \textbf{(a)} $Re = 800$, \textbf{(b)} $Re = 1600$, and \textbf{(c) $Re = 3000$}. The data-driven model (solid line) is compared against Smagorinsky (dashed line) and DNS benchmarks (symbols). The data-driven model exhibits superior accuracy in preserving kinetic energy during the transition phase compared to these classical models.}
    \label{fig:tgv_results}
\end{figure}

Figure \ref{fig:tgv_results} plots the temporal evolution of the resolved kinetic energy $E_k(t)$ and the dissipation rate $\mathcal{E}(t)$ across three different $Re$ ($800$, $1600$, and $3000$). While the visual structures appeared similar, the statistical results reveal significant differences in accuracy. As shown in Fig.~\ref{fig:tgv_re800}-\ref{fig:tgv_re3000}, the $\Phi$-SO model aligns almost perfectly with the DNS benchmarks \cite{Brachet_Meiron_Orszag_Nickel_Morf_Frisch_1983} across all tested Reynolds numbers. The evolution of $\mathcal{E}(t)$ further highlights the superiority of the data-driven approach. The peak of $\mathcal{E}(t)$ represents the moment of maximum non-linear interaction. The $\Phi$-SO model accurately predicts both the timing and the amplitude of this peak. The Smagorinsky model, however, under-predicts the peak dissipation rate, indicating that it overly dampens the intensity of the turbulent fluctuations during the critical transition phase.

\subsection{Validation on Inhomogeneous Shear Flow}
While the TGV case validates the model in decaying homogeneous turbulence, real-world flows typically involve wall boundaries and strong mean shear. To evaluate the model's performance in these more complex regimes, we examine the LDC flow at $Re=3200$. This configuration is characterized by a primary central vortex driven by the moving lid, along with a series of secondary counter-rotating vortices in the corners.

\begin{figure}[t]
    \centering
    \begin{subfigure}[b]{0.48\linewidth}
        \centering
        \includegraphics[width=\linewidth]{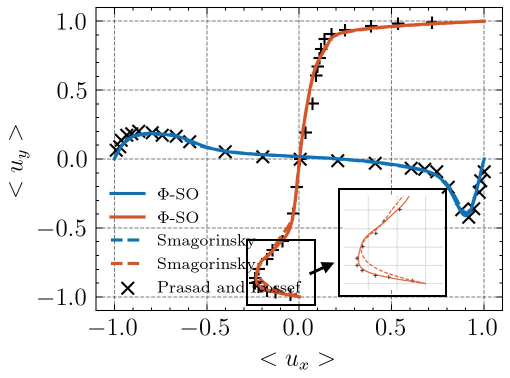}
        \caption{Mean velocity profile at $Re=3200$}
        \label{fig:ldc_mean}
    \end{subfigure}
    \hfill
    \begin{subfigure}[b]{0.48\linewidth}
        \centering
        \includegraphics[width=\linewidth]{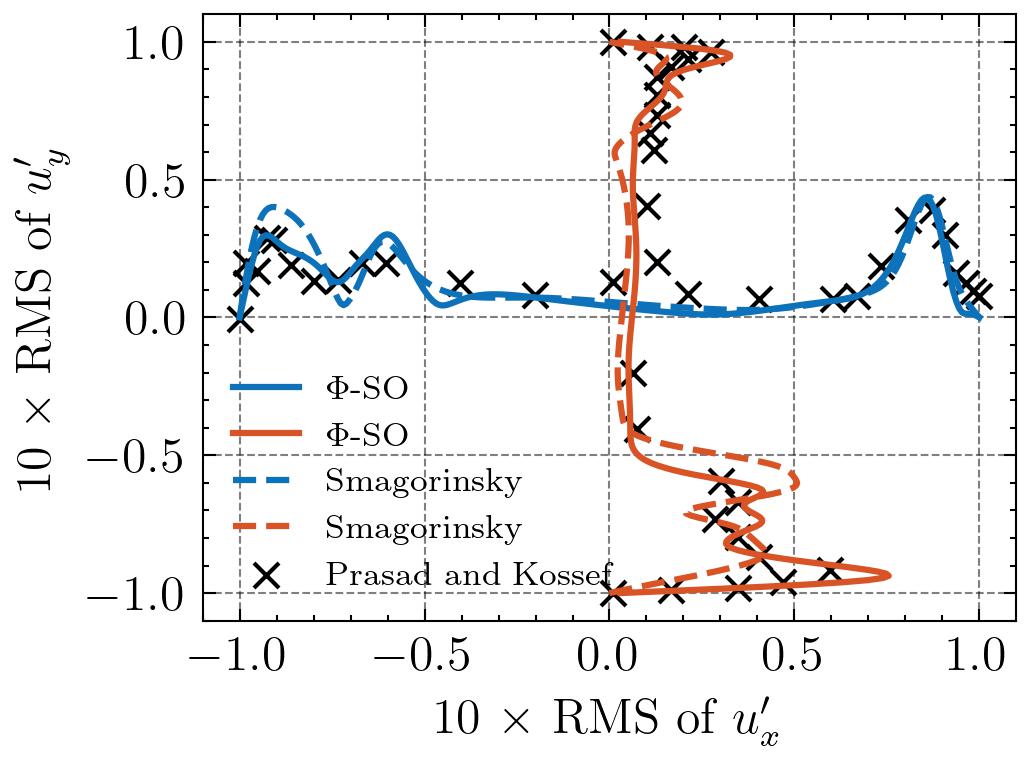}
        \caption{Fluctuation velocity at $Re=3200$}
        \label{fig:ldc_rms}
    \end{subfigure}
        \begin{subfigure}[b]{0.48\linewidth}
        \centering
        \includegraphics[width=\linewidth]{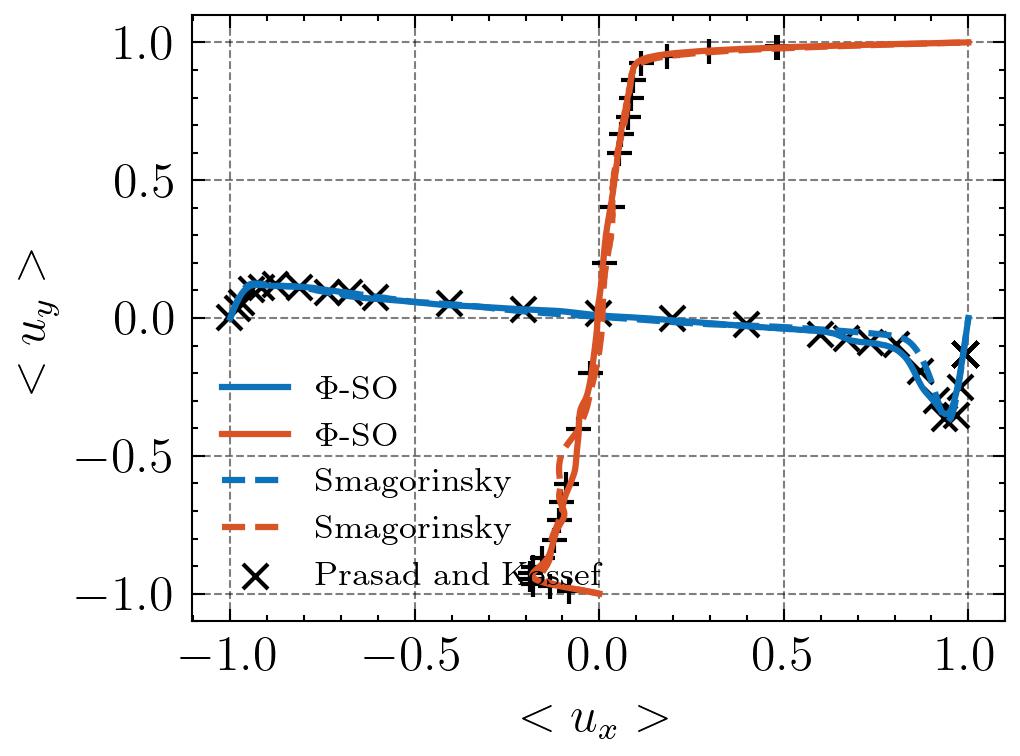}
        \caption{Mean velocity profile at $Re=10000$}
        \label{fig:ldc_mean_10000}
    \end{subfigure}
    \hfill
    \begin{subfigure}[b]{0.48\linewidth}
        \centering
        \includegraphics[width=\linewidth]{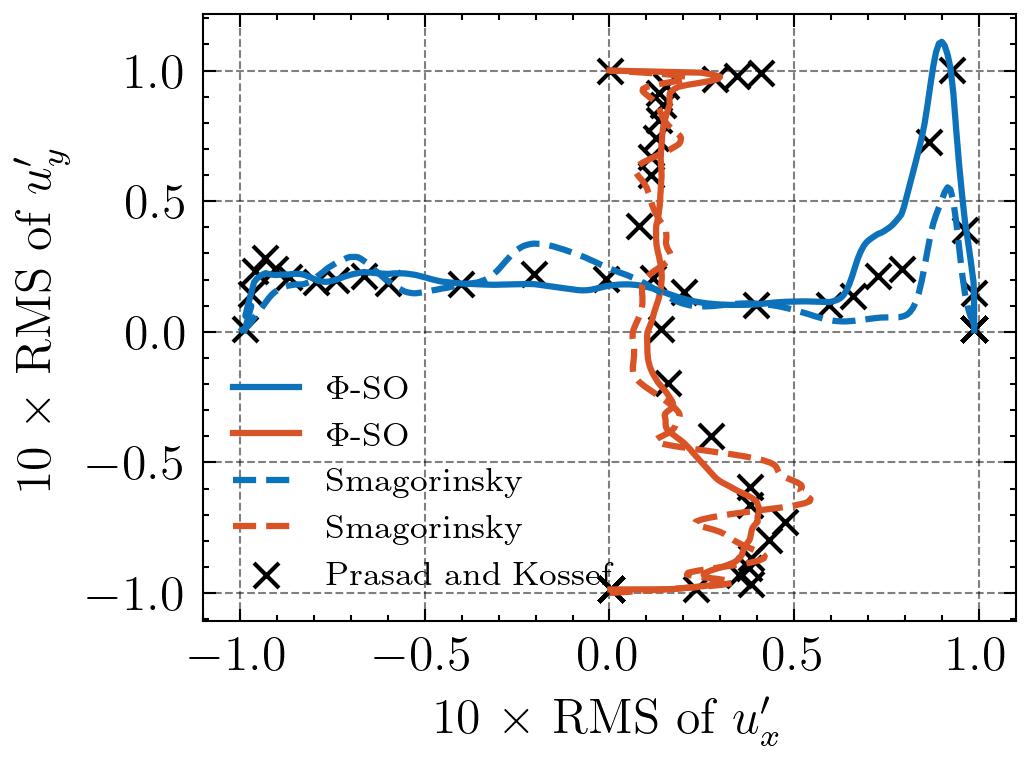}
        \caption{Fluctuation velocity at $Re=10000$}
        \label{fig:ldc_rms_10000}
    \end{subfigure}
    \caption{Velocity profiles for LDC flow at $Re = 3200$ and $Re = 10000$. Mean velocity profiles (a, c): horizontal velocity in red along the vertical centerline and vertical velocity in blue along the horizontal centerline. Fluctuation velocity profiles (b, d). The data-driven model (solid) closely matches the DNS reference data, particularly in the near-wall boundary layers, outperforming the classic Smagorinsky closure (dashed).}
    \label{fig:ldc_profiles}
\end{figure}

Figure \ref{fig:ldc_profiles} compares the velocity profiles obtained from the data-driven model against the DNS benchmarks \cite{PrasadAndKoseff} and the Smagorinsky model. Specifically, Fig.~\ref{fig:ldc_mean} and Fig.~\ref{fig:ldc_mean_10000} show the mean velocity profiles (horizontal velocity in red along the vertical centerline and vertical velocity in blue along the horizontal centerline) at $Re=3200$ and $Re=10000$, respectively. Fig.~\ref{fig:ldc_rms} and Fig.~\ref{fig:ldc_rms_10000} show the corresponding fluctuation velocity profiles. The results indicate that the $\Phi$-SO model achieves excellent agreement with the DNS benchmarks throughout the entire domain. The most significant improvement over the standard Smagorinsky model is observed in the near-wall regions and the high-shear boundary layers. The Smagorinsky model, due to its reliance on a constant coefficient, tends to overestimate the eddy viscosity in these high-shear zones, leading to worse results. In contrast, the symbolic model, which incorporates both strain and rotation invariants, dynamically adjusts the effective viscosity, thereby preserving the sharpness of the boundary layer profiles.

\begin{figure}[t]
    \centering
    \begin{subfigure}[b]{\linewidth}
        \centering
        \includegraphics[width=\linewidth]{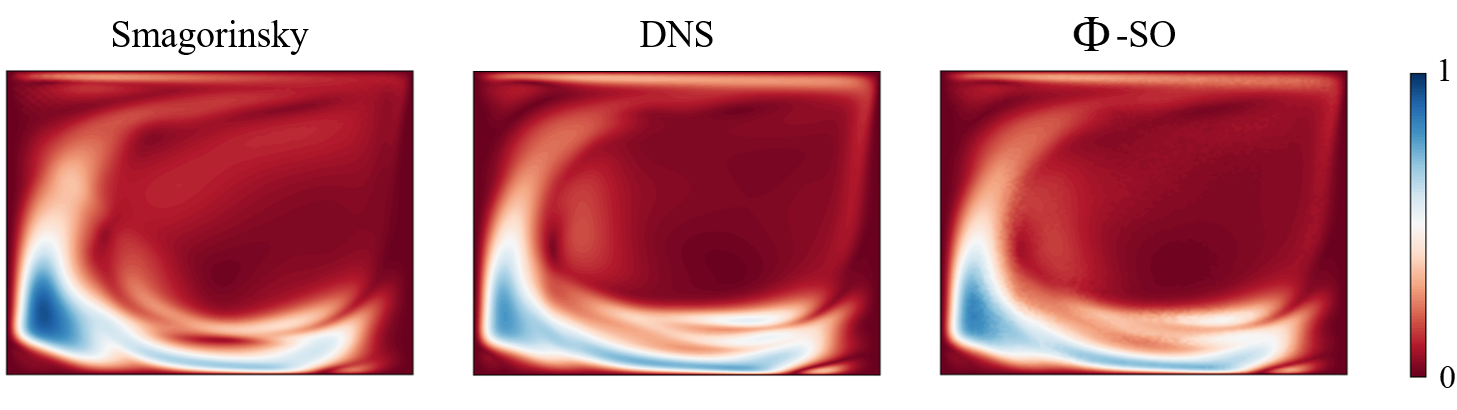}
        \caption{10 $\times$ root-mean-square (RMS) of horizontal fluctuation velocity ($u'_x$)}
    \end{subfigure}
    \begin{subfigure}[b]{\linewidth}
        \centering
        \includegraphics[width=\linewidth]{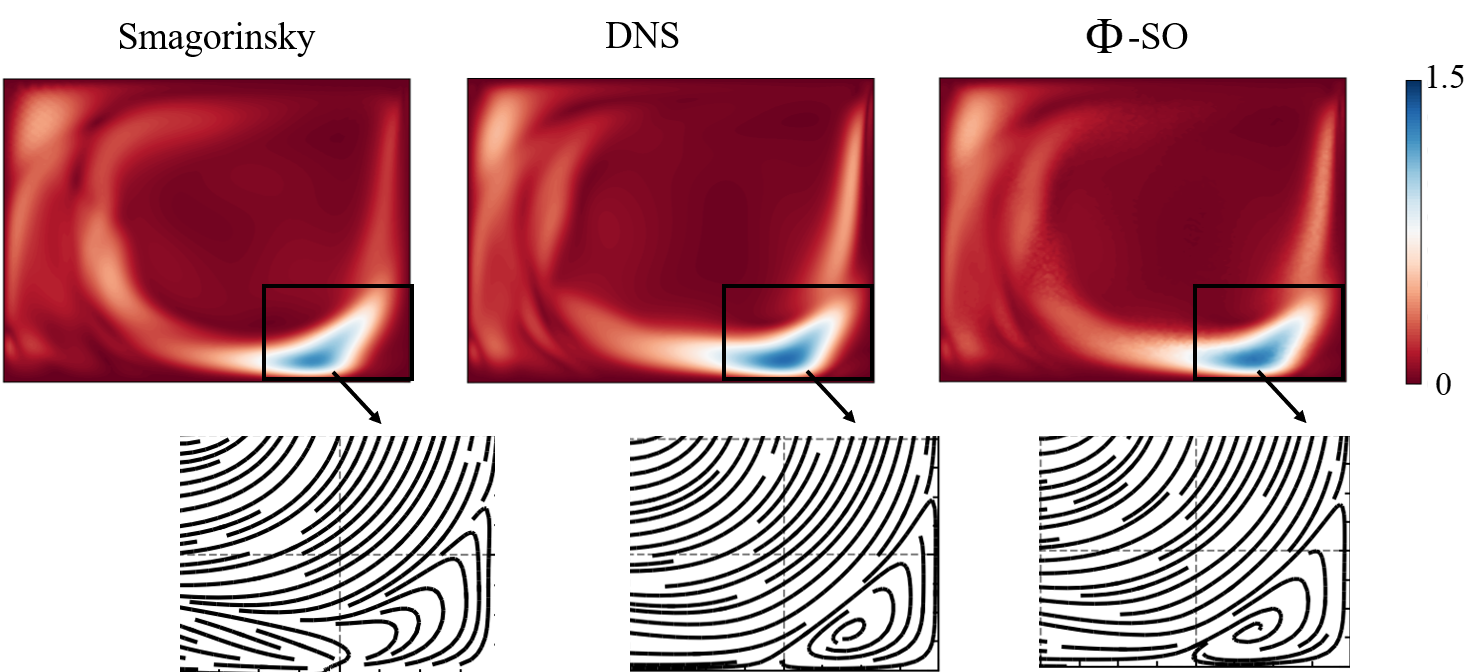}
        \caption{10 $\times$ RMS of vertical fluctuation velocity ($u'_y$) and secondary flow streamlines}
    \end{subfigure}
    
    \caption{Comparison of fluctuation velocity contours and streamlines for the LDC case. Note that the data-driven model ($\Phi$-SO) successfully captures the secondary corner vortices (Moffatt eddies), which are suppressed in the Smagorinsky simulation due to excessive eddy viscosity, matching the Ground Truth.}
    \label{fig:ldc_contours}
\end{figure}

Figure \ref{fig:ldc_contours} visualizes the RMS of fluctuation velocity contours and the corresponding streamlines in the bottom corners of the cavity.
These corner regions contain weak secondary recirculation zones known as Moffatt eddies. Capturing these structures is difficult for LES models because they exist in a delicate balance; excessive sub-grid dissipation can easily damp them out entirely.
As seen in Fig.~\ref{fig:ldc_contours}, the Smagorinsky model fails to sustain the corner vortices in both the horizontal and vertical fluctuation velocity fields, suppressing the recirculation due to excessive diffusion. In contrast, the $\Phi$-SO simulation successfully captures these fine-scale secondary vortices in both contour plots, reproducing the flow topology observed in the filtered DNS. This result demonstrates that the data-driven model correctly reduces the turbulent viscosity in laminar-like or weak-shear regions, overcoming a classic limitation of eddy-viscosity models.

\subsection{Generalization Test: Turbulent Channel Flow}
The ultimate test of a data-driven model is its ability to generalize to flow configurations outside its training distribution. While TGV and LDC represent decaying and recirculating flows respectively, the turbulent channel flow introduces a canonical wall-bounded regime dominated by stream-wise shear and near-wall anisotropy. Crucially, this geometry was not included in the training dataset, making this a true "blind test" of the model’s universality.

\begin{figure}[t]
    \centering
    \begin{subfigure}[b]{0.48\linewidth}
        \centering
        \includegraphics[width=\linewidth]{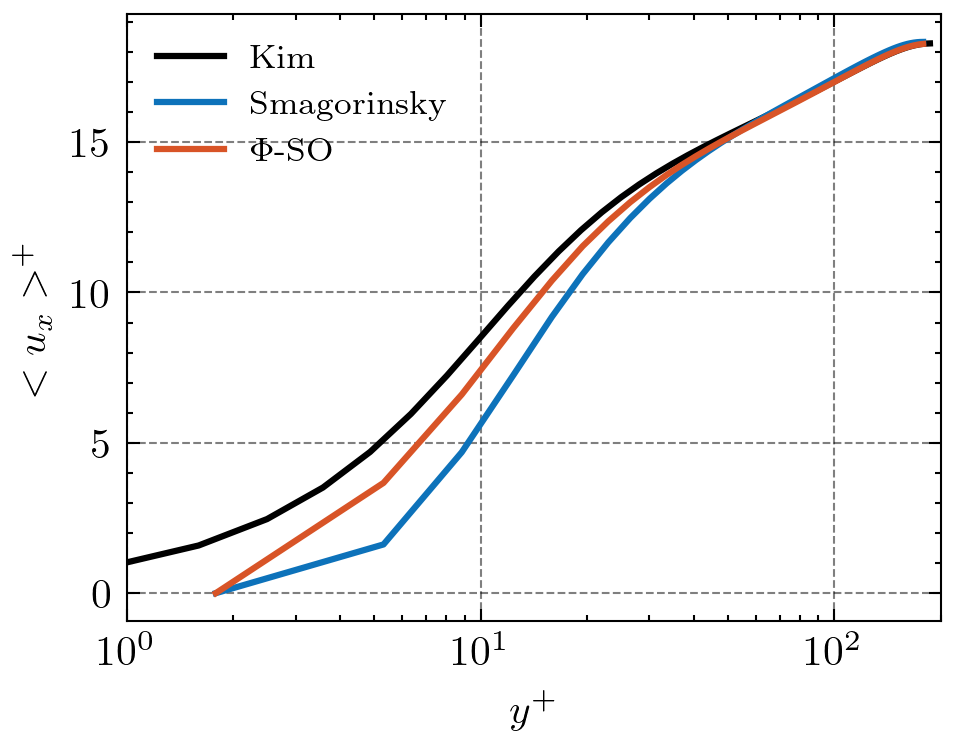}
        \caption{Mean Velocity ($u^+$ vs $y^+$)}
        \label{fig:channel_mean}
    \end{subfigure}
    \hfill
    \begin{subfigure}[b]{0.48\linewidth}
        \centering
        \includegraphics[width=\linewidth]{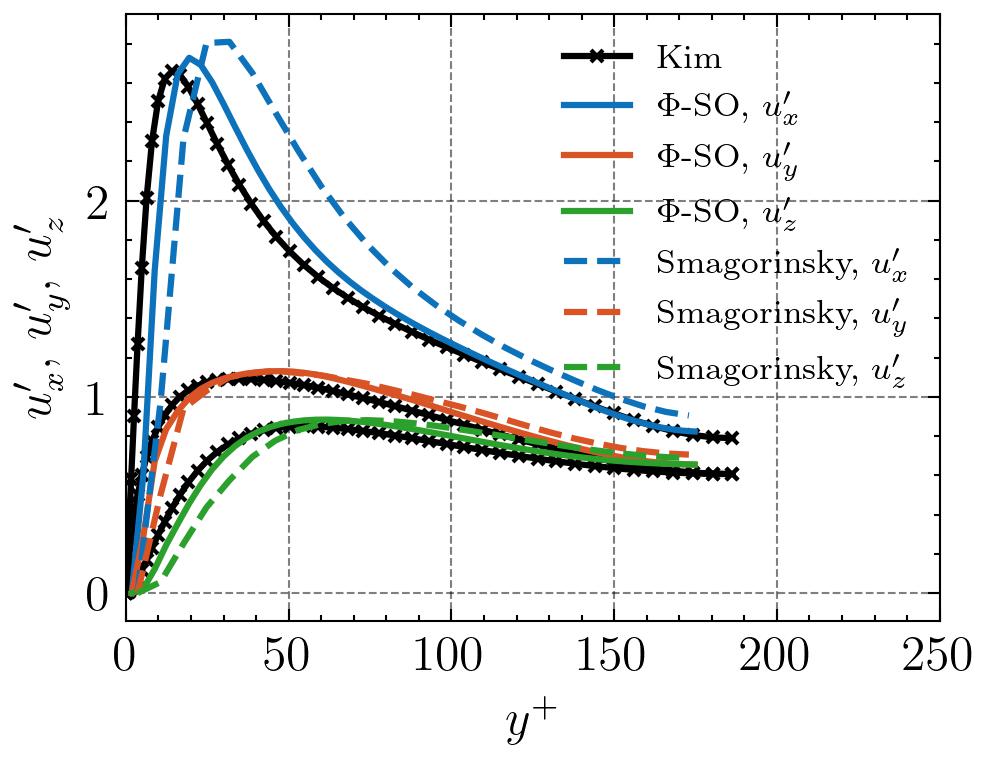}
        \caption{Velocity fluctuations}
        \label{fig:channel_rms}
    \end{subfigure}
    
    \caption{Generalization test on turbulent channel flow at $Re_{\tau} = 180$. \textbf{(a)} Mean streamwise velocity profile $u^+$ in wall units. The data-driven model (solid line) correctly reproduces the logarithmic law of the wall ($y^+ > 30$), showing broad agreement with Smagorinsky (dashed line) and DNS (symbols). \textbf{(b)} Velocity fluctuations. The data-driven model captures peak turbulence intensity near the wall ($y^+ \approx 20$) significantly better than Smagorinsky, demonstrating generalization to wall-bounded flows.}
    \label{fig:channel_results}
\end{figure}

We simulated a turbulent channel flow at a friction Reynolds number of $Re_{\tau} = 180$. The results are compared against the standard DNS database of Kim et al. \cite{kim1987}, as well as the classic Smagorinsky model. Figure \ref{fig:channel_results}(a) presents the mean streamwise velocity profile $u^+$ as a function of the wall-normal distance $y^+$ in semi-logarithmic coordinates. It is observed that both the $\Phi$-SO and Smagorinsky models show slight deviations from the DNS data in the immediate vicinity of the wall. This is attributed to the grid resolution limits of the LES, where the first lattice node is located at $y^+ \approx 1.37$. Despite this resolution constraint common in LES, the $\Phi$-SO model maintains a reasonable agreement and does not exhibit non-physical behavior. In the logarithmic region, the $\Phi$-SO model adheres more closely to the DNS profile, correctly reproducing the slope of the log law. This indicates that even without specific tuning for wall-bounded flows, the data-driven closure captures the essential physics of the inertial sublayer better than the standard model. 

Figure \ref{fig:channel_results}(b) displays the velocity fluctuations. As observed in the results, the Smagorinsky model overestimates the peak of the streamwise fluctuations ($u_x'$) while typically underestimating the transverse and spanwise components ($u_y'$ and $u_z'$). This indicates an inability to sufficiently redistribute energy from the streamwise mean flow to the other fluctuating components.
The data-driven model predicts a peak magnitude for the streamwise fluctuations that aligns more closely with the DNS data than the Smagorinsky model. This suggests that the symbolic formulation—likely through its inclusion of rotation and non-linear strain invariants—better captures the anisotropic mechanisms, preventing the excessive accumulation of energy in the streamwise component.

\section{Conclusion}
\label{sec:conclusion}
By integrating high-fidelity DNS data with the $\Phi$-SO algorithm, we successfully derived an explicit algebraic formulation for the effective relaxation time, $\tau_{\mathrm{eff}} = \tau_{\mathrm{turb}} + \tau_0$, effectively bridging the gap between data accuracy and physical interpretability.

The key findings and contributions of this work are summarized as follows:

\begin{itemize}
    \item \textbf{Physics-Informed Discovery Workflow:} We demonstrated that standard symbolic regression is significantly enhanced by embedding prior physical knowledge. The introduction of Virtual Dimensional Analysis and the construction of non-linear tensor invariants were critical. These constraints enabled the algorithm to navigate the vast search space effectively, identifying a parsimonious model that respects fundamental physical scalings (e.g., $\tau \propto |S|$).
    
    \item \textbf{Superiority in Transitional and Shear Flows:} A posteriori validation on the training configurations (TGV and LDC) confirmed the model's accuracy. In decaying isotropic turbulence, the data-driven model correctly predicted the kinetic energy dissipation rate during the critical transition phase, avoiding the excessive dissipation characteristic of the standard Smagorinsky model. In inhomogeneous shear flow, the model exhibited excellent spatial selectivity, remarkably capturing weak secondary corner vortices (Moffatt eddies) that are typically suppressed by conventional eddy-viscosity models due to over-diffusivity.
    
    \item \textbf{Robust Generalization to Wall-Bounded Turbulence:} The most significant result lies in the model's ability to generalize to Turbulent Channel Flow, a configuration absent from the training dataset. Even without ad-hoc near-wall damping functions, the model successfully reproduced the logarithmic law and maintained reasonable accuracy in the viscous sublayer, outperforming the baseline Smagorinsky model in capturing near-wall turbulent structures.
\end{itemize}

While the proposed model demonstrates promising potential, several avenues remain for future research. First, the current coefficients and fractional exponents are determined empirically; a rigorous theoretical derivation linking these terms to first-principles turbulence theory remains an open challenge. Second, assessing the model's stability and accuracy in high-Reynolds-number engineering applications is a necessary next step. Third, the model formulation contains a ratio $II_{\Omega}/II_{S}$ in Eq.~\eqref{eq:B}, which may lead to numerical instability in regions where $II_{S}$ approaches zero (e.g., in pure solid-body rotation); future work should address this potential singularity through appropriate regularization or reformulation. Fourth, comparing the proposed model against other advanced SGS models (e.g., WALE, Vreman, $\sigma$-model) would provide a more comprehensive evaluation of its performance. Ultimately, this work highlights the potential of symbolic regression not merely as a fitting tool, but as a mechanism for uncovering robust, interpretable physical laws, paving the way for the next generation of intelligent CFD solvers.
\section*{Statement}
During the preparation of this work, the authors used \emph{Gemini 3} in order to improve readability and language. After using this tool/service, the authors reviewed and edited the content as needed and take full responsibility for the content of the publication.
\section*{Acknowledgments}
This research was supported by the Jiangxi Provincial Natural Science Foundation (No. S20254688), Postdoctoral Fellowship Program of CPSF (No. GZC20250176), and Discretionary Fund of National Science Center for Earthquake Engineering (No. 2025ZZB4007).


%
%

%


\bibliography{cas-refs}
\end{document}